\documentclass[twocolumn,showpacs,preprintnumbers,prl,amsmath,amssymb]{revtex4}

\usepackage{graphicx}
\usepackage{dcolumn}
\usepackage{bm}
\usepackage{amsmath}

\begin{document}

\title{Probe-Configuration-Dependent Decoherence in an Aharonov-Bohm
Ring}

\author{Kensuke Kobayashi, Hisashi Aikawa} 
\author{Shingo Katsumoto$^\dagger$}
\author{Yasuhiro Iye$^\dagger$}

\affiliation{ Institute for Solid State Physics, University of Tokyo,
5-1-5 Kashiwanoha, Chiba 277-8581, Japan\\ $^\dagger$Also at CREST,
Japan Science and Technology Corporation, Mejiro, Tokyo 171-0031,
Japan }

\date{\today}

\begin{abstract}
We have measured transport through mesoscopic Aharonov-Bohm (AB) rings
with two different four-terminal configurations. While the amplitude
and the phase of the AB oscillations are well explained within the
framework of the Landaur-B\"uttiker formalism, it is found that the
probe configuration strongly affects the coherence time of the
electrons, i.e., the decoherence is much reduced in the configuration
of so-called nonlocal resistance. This result should provide an
important clue in clarifying the mechanism of quantum decoherence in
solids.
\end{abstract}



\maketitle

In mesoscopic physics, it is well known that probes for transport
measurements greatly affect the electronic states of the samples.
Particularly in single-electron phenomena, a low-impedance probe
introduces coupling of the system to an electromagnetic environment
and causes the degradation or even disappearance of the
single-electron charging effect~\cite{WatanabePRL2001}.  In coherent
transport, where probes and samples are strongly coupled, it was also
recognized in early studies that ``sample" and ``leads" are hardly
distinguishable~\cite{UmbachAPL1987}.

In the coherent regime, however, unlike the case of single-electron
phenomena, the quantum coherence of electrons has been believed to be
independent of the probe configuration, that is, the selection of the
pairs of voltage and current probes in a four-terminal
measurement. This assumption is likely to be true for the diffusive
transport regime because the electrons experience many elastic
scatterings in traversing the sample from one probe to another, which
would work as buffers to the environment.  Actually, it has been
verified both theoretically and experimentally that the phase
coherence time ($\tau_\phi$) in this regime is the same regardless of
their probe configurations~\cite{HauckePRB1990}.  On the other hand,
in ballistic transport, electrons go into samples directly from the
probes, making the boundaries more obscure.  In such circumstances,
the probes would affect the quantum coherence in the sample by
``observation" of electrons through the probes.

Quantum decoherence in solids is now one of the most important issues
in physics.  Mohanty \textit{et al.} examined the temperature
dependence of $\tau_\phi$ in various references and pointed out that
saturation of $\tau_\phi$ at the lowest temperatures is observed in
all cases~\cite{MohantyPRL1997}.  Subsequent studies have revealed
that this saturation is dependent not only on the
materials~\cite{GougamJLTP2000} but also on the sample
geometry~\cite{NatelsonPRL2001}, even in the diffusive regime.  These
suggest a possibility that the saturation is due to some extrinsic
effects.

In this Letter, we show that in the ballistic regime the coherence of
electrons greatly depends on the probe configuration, through a
four-terminal measurement of the \textit{same} sample with two
\textit{different} probe configurations.  First, we show that the
Landauer-B\"uttiker (LB) formalism, which was constructed for coherent
transport with all the probes treated on an equal
footing~\cite{ButtikerPRL1986}, successfully describes the qualitative
features of the present experiment. Next, this analysis is shown to
lead to an inevitable conclusion that the decoherence is dependent on
the probe configuration.

\begin{figure}[!tb]
\includegraphics[width=0.95\linewidth]{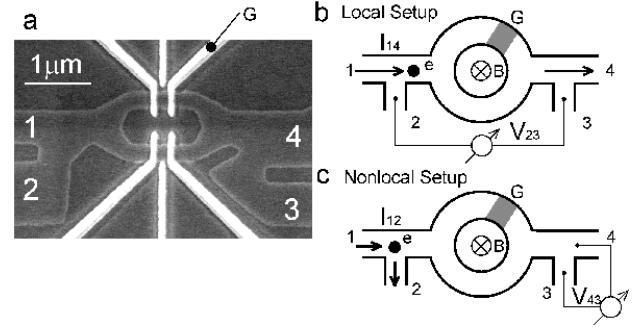}
\caption{\label{SampleFig} (a) Scanning electron micrograph of the AB
ring. One specific Au/Ti metallic gate marked ``G" was used in this
study. (b) Local four-terminal setup (current: 1$\rightarrow$4,
voltage: 2$\rightarrow$3).  (c) Nonlocal four-terminal setup (current:
1$\rightarrow$2, voltage: 4$\rightarrow$3). }
\end{figure}

We prepared an AB ring shown in Fig.~\ref{SampleFig} (a) in a
two-dimensional electron gas (2DEG) at an AlGaAs/GaAs heterostructure
(mobility of 9$\times$10$^5$~ cm$^2$/Vs and sheet carrier density of
3.8$\times$10$^{11}$~cm$^{-2}$) by electron-beam lithography and wet
etching.  The Fermi wavelength $2\pi/k_{\rm F}$ is estimated to be
40~nm. The electron mean free path $l_e\sim 8$~$\mu$m is longer than
the arms of the ring $L\sim 2$~$\mu$m, ensuring that the motion of
electrons in the ring is quasi-ballistic.  The Au/Ti metallic gates
with the width of $W\sim 100$~nm were deposited on the ring to control
the AB phase and one specific gate marked ``G" in Fig.~\ref{SampleFig}
(a) was used, with the others being kept open in the present study.

In AB oscillation measurements, electric current is usually passed
across the ring and the voltage drop between probes positioned along
the current direction is measured, as illustrated in
Fig.~\ref{SampleFig} (b).  Henceforth, we use the notation $R_{ij,kl}$
for the ``resistance" measured with the probe pair $(i,j)$ as current
probes and the pair $(k,l)$ as voltage probes.  Thus, the resistance
measured in the configuration in Fig.~\ref{SampleFig} (b) is denoted
as $R_{14,23}$ and is referred to as the ``local" setup.  Another
probe configuration possible for the same sample is the so-called
transfer-resistance configuration shown in Fig.~\ref{SampleFig} (c),
where the current flow is localized at one end of the ring and the
voltage appearing in the probe pair at the opposite end is measured
($R_{12,43}$ and ``nonlocal" setup).

In order to minimize the difference due to the instrumental effect,
the same setup and parameters were adopted for both configurations
except the current amplitudes, which were 3~nA and 15~nA for the local
and nonlocal setups, respectively.  It was confirmed that these
currents do not cause a Joule heating effect.  The resistances were
measured by the standard lock-in technique. A cryogenic low-pass
filter was inserted into every lead at the 1~K stage. Two samples
({\#}1 and {\#}2) with almost the same geometry were measured,
yielding essentially the same results.

\begin{figure}[!tb]
\includegraphics[width=0.95\linewidth]{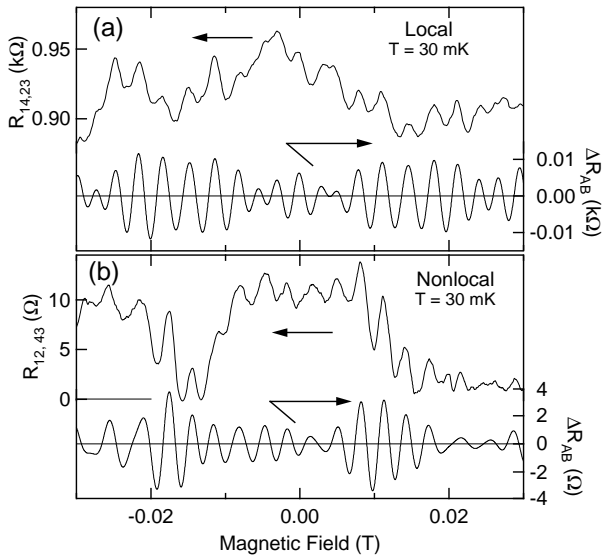}
\caption{\label{AB30mKFig} (a) The upper curve is a typical trace of
$R_{14,23}$ in the local setup. The lower curve shows the AB
component $\varDelta R_{\rm AB}$ extracted from $R_{14,23}$ through
FFT filtering.  (b) Corresponding data for the nonlocal setup.}
\end{figure}

The curves shown in the upper parts of Figs.~\ref{AB30mKFig} (a) and
(b) are the raw magnetoresistance traces for the local and nonlocal
setups, respectively.  The AB resistance oscillations with a period
$\varDelta B_{\rm AB}= 3.1\pm0.5$~mT, consistent with the ring size,
are superposed on the slower aperiodic fluctuation which arises from
small conductance loops within branches. The lower parts of
Figs.~\ref{AB30mKFig} (a) and (b) present the AB component ($\varDelta
R_{\rm AB}$) extracted through fast Fourier transform (FFT).  In the
local setup, $\varDelta R_{\rm AB}$ is only approximately 2~\% of the
total resistance, which is a typical value for the oscillation
amplitude seen in many previous AB experiments. By contrast, in the
nonlocal setup, it averages up to 20~\% with a maximum of $\sim
75$~\%.  This is the largest AB amplitude ever reported.

\begin{figure}[!tb]
\includegraphics[width=0.95\linewidth]{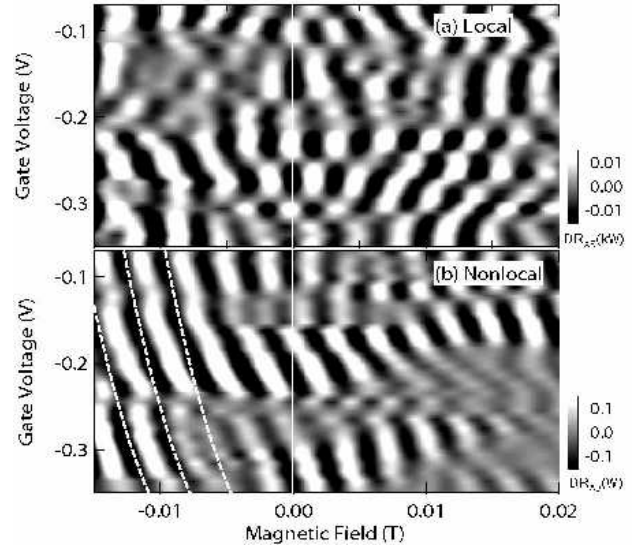}
\caption{\label{ABphase} Gray-scale plots of the AB oscillation
components of the resistances as a function of $V_{\rm G}$ observed in
(a) the local setup and (b) the nonlocal setup.  In (a), the AB
oscillation is almost symmetric with respect to $B=0$ (white vertical
line).  The white dashed curves in (b) are expected equiphase lines in
the $B$-$V_{\rm G}$ plane. }
\end{figure}

We also performed electrostatic control of the electronic phase
through the voltage of the gate G ($V_{\rm G}$).  The gray-scale plots
in Fig.~\ref{ABphase} show the gate-voltage-induced variation of the
AB oscillation at 30~mK.  The variation of the electrostatic potential
energy by $V_{\rm G}$ results in that of the kinetic energy, and hence
the wave number of the electrons which traverse the region underneath
the gate electrode.

Now we demonstrate that the qualitative features of the above results
at lowest temperatures are well explained within the LB framework.
The LB formula gives
$R_{mn,kl}=(h/2e^2)(T_{km}T_{ln}-T_{kn}T_{lm})/D$, where
$T_{ij}$~$(\ge 0)$ is the transmission coefficient from terminal $j$
to $i$ and the denominator $D$ is a quantity including all of the
$T_{ij}$'s~\cite{ButtikerPRL1986}.  Here, we note that this formula,
which was originally applied to perfect coherent
transport~\cite{ButtikerPRL1986}, can be generalized to include
incoherent transport~\cite{ButtikerIJRD}. The generalized LB formula
is expressed by the transmission coefficients that are the sum of a
coherent part and an incoherent part, and has the same functional form
as the original one. Thus the effect of decoherence is appropriately
taken into account through the amplitude of the interference term in
the transmission coefficients in the LB formula.

From the sample geometry shown in Fig.~\ref{SampleFig} (a), it is
natural to make the approximation that $T_{12}=T_{21}=T_{34}
=T_{43}\equiv T_0$, $T_{14}=T_{41}=T_{23}=T_{32}\equiv T_1$ and
$T_{13}=T_{31}=T_{24}=T_{42}\equiv T_2$. In order to take the
decoherence into account, transmission coefficients for those paths
traversing the ring ($T_1$ and $T_2$ in the present case) are taken as
$T_{ij}=\alpha_{ij}+\beta_{ij}\cos(2\pi\phi/\phi_0+\delta_{ij})$,
where $\beta_{ij}$ represents the AB amplitude, $\alpha_{ij}$ the
other part of the transmission, $\phi$ the magnetic flux which threads
the ring, and $\phi_0\equiv h/e$ the flux quantum.  $\delta_{ij}$ is
the phase due to the difference between the paths.  Although
$\alpha_{ij}$ also depends on the magnetic field and contains the
terms due to the interference,
$|\overline{\beta_{ij}}/\overline{\alpha_{ij}}|$ serves as a good
measure of the coherence of traversing electrons, where the bars
denote the average over a certain range of magnetic field.

In the above approximation,  the LB formula gives the resistances
\begin{gather}
R_{14,23}=\frac{h}{2e^2}\frac{T_0-T_2}{(T_0+T_1)(T_1+T_2)}
\sim\frac{h}{4e^2}\frac{1}{T_1+T_2},\label{local}\\
R_{12,43}=\frac{h}{2e^2}\frac{T_1-T_2}{(T_0+T_1)(T_0+T_2)}
\sim\frac{h}{4e^2}\frac{T_1-T_2}{T_0^2},\label{nonlocal}
\end{gather}
where we further approximate that $T_0\gg T_1,T_2$.  Note that the
information on the phase of the AB oscillation has been neglected in
this approximation because of the symmetric assumption
$T_{ij}=T_{ji}$~\cite{ButtikerIJRD}.  Equation~\eqref{local} indicates
that the ratio of the AB oscillation in $R_{14,23}$ directly reflects
$\overline{\beta}/\overline{\alpha}$ while in Eqn.~\eqref{nonlocal}
the incoherent parts in $T_1$ and $T_2$ cancel each other by
subtraction and the AB amplitude is emphasized in the
magnetoresistance.  This is the qualitative explanation for the
results shown in Fig.~\ref{AB30mKFig}.

Next, the phase modulation by the gate voltage $V_{\rm G}$ in
Fig.~\ref{ABphase} is explained as follows.  The applied gate voltage
modulates the potential of electrons and hence their wave numbers,
which results in the variation of $\delta_{ij}$.  According to a
capacitance model between the gate and the 2DEG~\cite{YacobyPRL1991},
the equi-phase curve of the AB oscillation in the $B$-$V_{\rm G}$
plane should take the form $B=\varDelta B_{AB}(Wk_{\rm
F}/2\pi(1-|V_{\rm G}/V_C|)^{1/2}+N)$ $(N=0,\pm 1,\dots)$, where $V_C$
is the pinch-off voltage ($-0.435$ V for the present case). This curve
satisfactorily explains the overall trend of the phase shift in the
nonlocal case as indicated by the white dashed lines in
Fig.~\ref{ABphase} (b) (the curves for $N=-1, 0$, and $1$ are
plotted).  The phase jumps observed in a few regions are due to the
intermixing effect of the conduction
channels~\cite{CernicchiaroPRL1997}.

On the other hand, the phase variation in the local case differs
greatly from the above.  It was established by early studies that in
the LB framework a two-terminal resistance $R_{tt}$ must be an even
function of the magnetic field, i.e., $R_{tt}(B)=R_{tt}(-B)$ due to
Onsager's reciprocal theorem $T_{ij}(B)=T_{ji}(-B)$ and current
conservation~\cite{ButtikerIJRD}.  The reciprocal theorem appears in
four terminal resistances as $R_{kl,mn}(B)=R_{mn,kl}(-B)$ and
$R_{kl,mn}$ itself is not necessarily even with $B$.  Hence, it is
natural that the phase varies continuously with $V_{\rm G}$ in the
nonlocal setup.  Actually, we experimentally confirmed that
$R_{12,43}(B)=R_{43,12}(-B)$ holds.  In the local setup, the current
and voltage probes are placed crossing the ring part, which brings the
situation close to the two-terminal case and imposes the symmetry for
the magnetic-field reversal.  This results in the symmetric pattern
with respect to $B=0$ as shown in Fig.~\ref{ABphase} (a).  As reported
previously~\cite{CernicchiaroPRL1997}, the phase jump by $\pi$ occurs
at several values of $V_{\rm G}$. Also, the fluctuation of the AB
period due to the mixing of conduction channels causes the
quasi-continuous shift of the oscillation though the symmetry for
$B=0$ still holds.

\begin{figure}[!tb]
\includegraphics[width=0.97\linewidth]{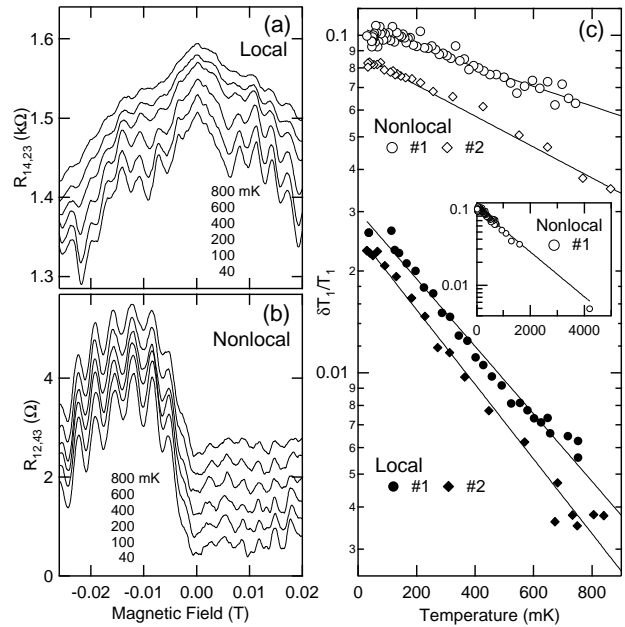}
\caption{\label{ABTdepFig} (a) AB oscillation for the local setup
measured at various temperatures.  (b) Counterpart of (a) for the
nonlocal measurement. The data at $T\geq$~100~mK are incrementally
shifted upward by 0.02~k$\Omega$ in (a) and 0.4~$\Omega$ in (b).  The
temperature dependence of the AB effect is clearly different between
(a) and (b).  (c) Temperature dependence of the portion of the AB
amplitude in the transmission coefficient ($\delta T_1/T_1$) for the
two configurations and two different samples ({\#}1 and {\#}2) with
the same geometry.  The solid lines are the results of the fitting to
the exponential decay of temperature. The inset shows the nonlocal
result up to 4.2~K. }
\end{figure}

Thus far, we have seen that the LB formula successfully describes the
qualitative profiles of the results, which leads us to believe that
the quantitative outcome of the formula can be used as a basis for the
comparison between the two configurations. In order to estimate the
classical $T_{ij}$'s, we measured the resistances for all of the
possible probe configurations at 4.2~K where the quantum interference
effect is small.  By combining these values with the AB oscillation
measured at various temperatures presented in Figs.~\ref{ABTdepFig}
(a) and (b), we obtain the temperature dependence of the portion of
the interference term in the transmission coefficient $\delta T_1/T_1$
as shown in Fig.~\ref{ABTdepFig} (c). A striking difference between
the local and the nonlocal setup is observed: The ratio of the
coherent term to the total transmittance is much higher in the
nonlocal probe configuration and it is much more robust against
thermal disturbance.  As shown in Fig.~\ref{ABTdepFig} (c), the
consistency of results between the two different samples with the same
geometry strongly indicates that this observation is not
sample-specific but is due to the measurement configuration.

Two factors are important for the reduction of the AB amplitude with
temperature. One is the thermal broadening of the electron wave
packets, which is expressed in the ballistic case as
$\beta_{ij}\propto\exp(-\tau_Lk_{\rm B}T/\hbar) \equiv
\exp(-\tau_L/\tau_{\rm th})$, where $\tau_L=L/v_{\rm F}$ ($v_{\rm F}$:
the Fermi velocity).  The other is the quantum decoherence, which
contributes as
$\beta_{ij}\propto\exp(-\tau_L/\tau_\phi(T))$~\cite{SeeligPRB2001}.
As shown by the solid lines in Fig.~\ref{ABTdepFig} (c), the present
data indicate $\beta\propto \exp(-aT)$, implying that $\tau_\phi$ is
negligible or is also proportional to $T^{-1}$. The obtained values of
$a$ are 0.72 and 1.0~K$^{-1}$ for the nonlocal results, while $a=2.3$
and 2.5~K$^{-1}$ for the local ones.  The thermal broadening affects
as $\exp(-\tau_L/\tau_{\rm th}) \equiv \exp(-bT)$ with $b \sim
1$~K$^{-1}$ in the present case, which would be sufficient to explain
the temperature dependence of the nonlocal result. Since the thermal
broadening is an intrinsic effect, we attribute the observed
probe-configuration-dependent effect to the difference in $\tau_\phi$.

Then the next problem is the origin of the difference in $\tau_\phi$
between the two setups, or rather the suppression of $\tau_\phi$ in
the nonlocal setup. Here, we would like to note that the electric
current flows differently in that electrons run quasi-ballistically
between terminals 1 and 4 in the local setup, while no net current
enters the interferometer in the nonlocal setup.  In the LB method,
zero net current is the consequence of the cancellation of back and
forth currents through a sample, but these ``currents" are merely
means for the calculation and are not physical entities.  What
actually happens and what the LB formula actually describes in the
nonlocal setup is that an electrostatic charge up occurs between
electrodes 3 and 4, which prevents the current flow.

We list here two candidates for the origin. The first is that the
electrons in the local setup suffer extra decoherence due to the
charge fluctuation.  It is proposed that in a ballistic AB ring the
charge fluctuations caused by a nearby capacitor such as metallic
gates result in decoherence expressed as $\tau_\phi\propto
T^{-1}$~\cite{SeeligPRB2001}, which is in agreement with our
observation. A similar observation was reported by Hansen \textit{et
al.}~\cite{HansenPRB2001}.  In the local setup, charge fluctuation,
which originates from shot noise~\cite{ReznikovSLMS1998} due to the
nonequilibrium current in the interferometer, is possible to play a
significant role when the net current exists.  In the nonlocal setup,
this mechanism is expected to be suppressed.  The second mechanism is
based on the deduction that in the local setup injected electrons
traverse the ring quasi-ballistically and may retain some traces that
can tell ``which path they went through" to the circumambient
electrons in 2DEG~\cite{BuksNature1998}.  In other words, the
interference does not complete at the outlet of the sample locally.
This should lead to a loss of coherence through the process discussed
by Stern \textit{et al.}~\cite{SternPRA1990}.  In the nonlocal setup,
the electrons in the ring are comparatively isolated from the leads by
the electrostatic barrier formed by the charge up.  This mechanism may
be reduced by making the outlet narrower to prevent the leakage of the
``which path" information.

In conclusion, we have observed that the quantum decoherence in a
ballistic AB ring is dependent on the probe configuration.  While
$\tau_\phi$ behaves as $\propto T^{-1}$, in both local and nonlocal
setups, the coherence survives at much higher temperatures in the
nonlocal setup. We believe that the key difference is the amplitude of
the net electric current, and plausible mechanisms are discussed.
While theoretical effort is necessary to clarify this issue, the LB
formula successfully describes the qualitative profiles of the
results, such as the large amplitude of the AB oscillation in the
nonlocal setup or the phase tuning by electrostatic potential.

We thank H. Ebisawa and H. Fukuyama for helpful discussion.  This work
is partly supported by a Grant-in-Aid for Scientific Research and by a
Grant-in-Aid for COE Research (``Quantum Dot and Its Application")
from the Ministry of Education, Culture, Sports, Science, and
Technology.

\end{document}